
\magnification=1200
\baselineskip=20pt
\def\sqr#1#2{{\vcenter{\vbox{\hrule height.#2pt
        \hbox{\vrule width.#2pt height#1pt \kern#1pt
          \vrule width.#2pt}
        \hrule height.#2pt}}}}

\def\boxit#1{\vbox{\hrule\hbox{\vrule\kern3pt
   \vbox{\kern3pt#1\kern3pt}\kern3pt\vrule}\hrule}}
\def\lsim{<\kern-2.5ex\lower0.85ex\hbox{$\sim$}\ }
\def\rsim{>\kern-2.5ex\lower0.85ex\hbox{$\sim$}\ }
\overfullrule=0pt
\def\LAMBDABAR {\hbox{$\lambda$\kern-0.52em\raise+0.45ex\hbox{--}\kern+0.2em}}
\def\partialbar {\hbox{$\partial$\kern-0.52em\raise0.2ex\hbox{/}\kern+0.1em}}
\def\abar {\hbox{A\kern-0.6em\raise0.2ex\hbox{/}\kern+0.1em}}
\def\ebar {\hbox{E\kern-0.6em\raise0.2ex\hbox{/}\kern+0.1em}}
\centerline{\bf Comment on \lq\lq Operator Algebra in}
\centerline{\bf Chern-Simons Theory on a Torus"}
\vskip 1cm
\centerline{by}
\vskip 1cm
\centerline{C.R. Hagen}
\centerline{Department of Physics and Astronomy}
\centerline{University of Rochester}
\centerline{Rochester, NY 14627, USA}
\centerline{and}
\centerline{E.C.G. Sudarshan}
\centerline{Center for Particle Physics}
\centerline{Department of Physics}
\centerline{University of Texas at Austin}
\centerline{Austin, TX 78712}
\vfill\eject
Recent papers by Hosotani$^1$ $(H)$ and Ho and Hosotani$^2$ $(H^2)$ have
sought to give a dynamical meaning to certain time dependent variables
$\theta_i(t)$ (called \lq\lq nonintegrable phases of Wilson line integrals"
in $H$ and $H^2$) which occur in the solution of the pure Chern-Simons
theory$^3$ on a torus.  In an earlier Comment$^4$ it was shown that the
resulting quantization of the Chern-Simons coefficient found in $H$ for the
Abelian model was merely a consequence of errors made there in solving the
field equation
$${\kappa \over 2 \pi} \ \epsilon^{\mu \nu \alpha} \partial_\nu a_\alpha = -
J^\mu \ \ .\eqno(1)$$

The aim of this subsequent Comment is to indicate even more clearly the
lack of dynamical content in the $\theta_i (t)$ parameters.  To this end
the solution obtained in ref. 4 will be summarized (using $H^2$ notation),
the differences with $H$ and $H^2$ clearly noted, and the obvious validity
of our modifications pointed out.  The solution of (1) is from ref. 4
$$\eqalign{a_i &= \partial_i \Lambda + {2 \pi \over \kappa} \ \epsilon_{ij}
\nabla_j \int d^2y\  {\cal D} (x,y) J^0 (y)\cr
\noalign{\vskip 4pt}%
a_0 &= \partial_0 \Lambda + {2 \pi \over \kappa}
 \int d^2y \vec J \times \vec \nabla_y {\cal D}
(x,y) \cr} \eqno(2)$$
where
$\Lambda = \sum x_i \theta_i (t) /L_i$
and
$${\cal D} (x,y) = D (x-y) + {x^2 + y^2 \over
4L_1 L_2} = {\cal D} (y,x)$$
with
$$\nabla^2 {\cal D} (x,y) = \nabla^2 D (x-y) + {1 \over L_1 L_2}
= \delta (x-y) \ \ .$$

The defense$^5$ of the solution found in ref. 1 was threefold -- namely,
 (a) the assertion that a certain theorem excluded $x$-dependent terms from
the solution, (b) that ref. 4 improperly identified a $c$-number $q$ with
an operator $Q$, and (c) that the meaning of time derivative terms
$\dot{\theta}_i (t)$ is \lq\lq not clear".  In response it should be
pointed out that (a) since the Laplacian of a term linear in $x$
necessarily vanishes, the theorem of ref. 5 is incorrect; (b) Eq. (7) of
ref. 5 unambiguously identifies $a$ and thus $q$ with the charge operator; and
(c) the $\dot{\theta}_i(t)$ of ref. 4 have now also been used by $H^2$ (Eq.
(7)).

With regard to the more recent work $(H^2)$ there are three significant
points of difference between (2) and $H^2$.

\noindent i) $H^2$ deny $Q = -{\kappa \over 2 \pi} \Phi$ as an operator
relation.  However, insertion of their result$^6$ for $a_i$ into the temporal
component of (1) leads to an immediate verification of that result.  Note
that this also suffices  to establish the vanishing of $[P_i, P_j]$ and
 $[P_i, {\cal H}]$, in contrast to the claim of $H^2$.
 (The operators $P_i$ and
${\cal H}$ are the momentum and Hamiltonian operators  respectively.)

\noindent ii) $H^2$ do not have the term $\partial_0 \Lambda$ in $a_0$
which gives a contribution $\sum_i x_i \dot{\theta}_i/ L_i$
 to that operator.  This
additional term is clearly necessary to cancel the corresponding
$\dot{\theta}_i$ term in $\partial_0 a_i$ when one considers the spatial
components of Eq. (1).  It is also clearly compatible with the boundary
conditions (3) of $H^2$, the contrary claim$^5$ of Hosotani
notwithstanding.

\noindent iii) There is a contribution to $a_0$ arising from the difference
between ${\cal D}$ and $D$ which is absent in $H^2$.  This can readily be
seen to be required by the $[{\cal H}, \psi]$ result.  The point is
 that the $a \cdot J$ term in ${\cal H}$
 gives (as a consequence of a term proportional to
$\epsilon_{ij} x_j Q$ in $a_i$) a contribution to $\dot{\psi}$ which goes
as $ \int \vec x \times \vec J \ d^2x$.  This is accommodated within the
 form given by Eq. (2) for $a_0$.

These errors in $H$ and $H^2$ significantly affect
the role of the $\theta_i$'s.
Because there is no $\dot{\theta}$ term in their $a_0$ the cancellation
between $\dot{\theta}$ terms in the Lagrangian which is implied by (2)
simply does not occur.  Consequently nonvanishing commutators are found for the
$\theta_i$'s amongst themselves and nontrivial equations of motion are
 implied for these variables.

Thus $H$ and $H^2$ have incorrectly solved the
equations of motion for the pure Chern-Simons theory on a torus.  When the
appropriate corrections are made, the $\theta_i$'s
evidently become nondynamical quantities whose sole effect is that of a gauge
transformation on the true dynamical variables of the system.
 A study of the effects of such
functions on the Hilbert space (and associated issues) will
 not therefore accomplish
anything beyond what is already implied by the usual gauge invariance of
the Chern-Simons theory.$^7$
\vfill\eject
\noindent {\bf Acknowledgment}

This work is supported in part by the U.S. Department of Energy Grants
No. DE-FG-02-91ER40685 (CRH) and DE-FG-05-85ER40200 (ECGS).

\bigskip
\bigskip
\noindent {\bf References}

\item{1.} Y. Hosotani, Phys. Rev. Lett. {\bf 62}, 2785 (1989).
\item{2.} C-L Ho and X. Hosotani, Phys. Rev. Lett. {\bf 70}, 1360 (1993).
\item{3.} C. R. Hagen, Ann. Phys. (NY) {\bf 157}, 342 (1984).
\item{4.} C. R. Hagen and E. C. G. Sudarshan, Phys. Rev. Lett. {\bf 64},
1690 (1990).
\item{5.} Y. Hosotani, Phys. Rev. Lett. {\bf 64}, 1691 (1990).
\item{6.} It is not without interest to observe that this result has also
been derived in ref. 1.
\item{7.} If, however, $\theta_i(t)$ are added to $a_i$ but
\underbar{not} constrained by equations of motion (1) then nontrivial
vacuum structure and multicomponent wave functions obtain.  Examples are R.
Iengo and K. Lechner, Nucl. Phys. {\bf B364}, 551 (1991) and K. Lechner,
Phys. Lett. {\bf B273}, 463 (1991).  These structures are reproduced by
$H^2$.

\end